\documentstyle[prb,aps,twocolumn]{revtex}
\input epsf
\begin{document}
\draft

\title{Charge transport through quantum dots via time-varying tunnel couplings
}

\author{F. Renzoni $^1$, T. Brandes $^2$}

\address{$^1$ Laboratoire Kastler-Brossel, Departement de Physique de l'Ecole
Normale Sup\'erieure, 24 rue Lhomond, 75231, Paris Cedex 05, France}

\address{ $^2$ Department of Physics, University of Manchester Institute of
Science and Technology (UMIST), P.O. Box 88, Manchester M60 1QD, United 
Kingdom} 

\date{\today{ }}
\maketitle 
\begin{abstract}
We describe a novel mechanism for charge pumping through tunnel--coupled
quantum dots in the regime of strong Coulomb blockade. The quantum state 
of an additional electron within the structure is steered by changing the
tunneling couplings between neighbouring dots. Appropriate tailoring
of the interdot tunneling rates allows to design the instantaneous 
eigenvalues of the system Hamiltonian.
A combination of adiabatic following and Landau--Zener tunneling
results in the transfer of charge from one dot to the neighbouring one.
Coupling to electron reservoirs via weak tunnel--barriers then allows to
implement an electron pump.  
\end{abstract}


Single electron devices allow for the controlled transfer and pumping of 
charges through small metallic islands or semiconductor quantum dots 
\cite{pumpexperiment,Grabert}.  Monochromatic time--dependent electric 
fields have been demonstrated \cite{micro} to lead to photo--assisted
tunneling through coupled quantum dots \cite{stoof,sung}, where resonant 
tunneling via two controllable discrete levels can be modulated by applying 
an oscillating signal to the gate electrode or by irradiating the structure
with microwaves.

Consequently, different schemes for double dot pumps, i.e. devices  for the
transfer of electrons between two reservoirs at the same chemical potential,
have been suggested for monochromatic \cite{pumpdevices} and pulsed 
irradiation \cite{pulsed}. 

An alternative pumping mechanism is the slow parametric change of system 
parameters such as tunnel rates. Although the original single electron
turnstile experiments \cite{pumpexperiment} relied on adiabatic electron 
transfer, the adiabatic control of the wave function itself is a relatively
new topic. 
%
%
Experiments in open dots \cite{switkes} have demonstrated the feasability of 
an `adiabatic quantum electron pump'. 
These systems can be described as  non--interacting  mesoscopic scatterers
\cite{brouwer}. This allows for the generalization of a number of concepts
from metallic systems, such as mesoscopic fluctuations \cite{zhou,aleiner}, 
symmetries \cite{aleiner}, or resonances \cite{wei}, to the time--dependent 
case.

In this work, we describe a novel mechanism for charge pumping through 
{\em three} tunnel--coupled quantum dots  attached to electron leads 
(reservoirs) in the regime of strong Coulomb blockade. 
The main idea is to achieve pumping  by periodically varying not the coupling
to the leads but by varying the (inner) couplings among the dots, which are
an intrinsic part of the whole quantum system (the triple dot) itself. 
Appropriate tailoring of the tunneling rates between neighbouring dots then 
allows to design the instantaneous energies and wave functions of the structure.
A combination of adiabatic following and Landau--Zener tunneling 
results in the transfer of charge from one dot to the neighbouring one.
Coupling to electron reservoirs via constant tunnel--barriers then allows to 
implement an electron pump completely based on a quantum mechanical mechanism.

\begin{figure}[ht]
\setlength{\unitlength}{1in}
\begin{picture}(4,1)
\put(0.25,0){\epsfxsize 2.75 in \epsfbox{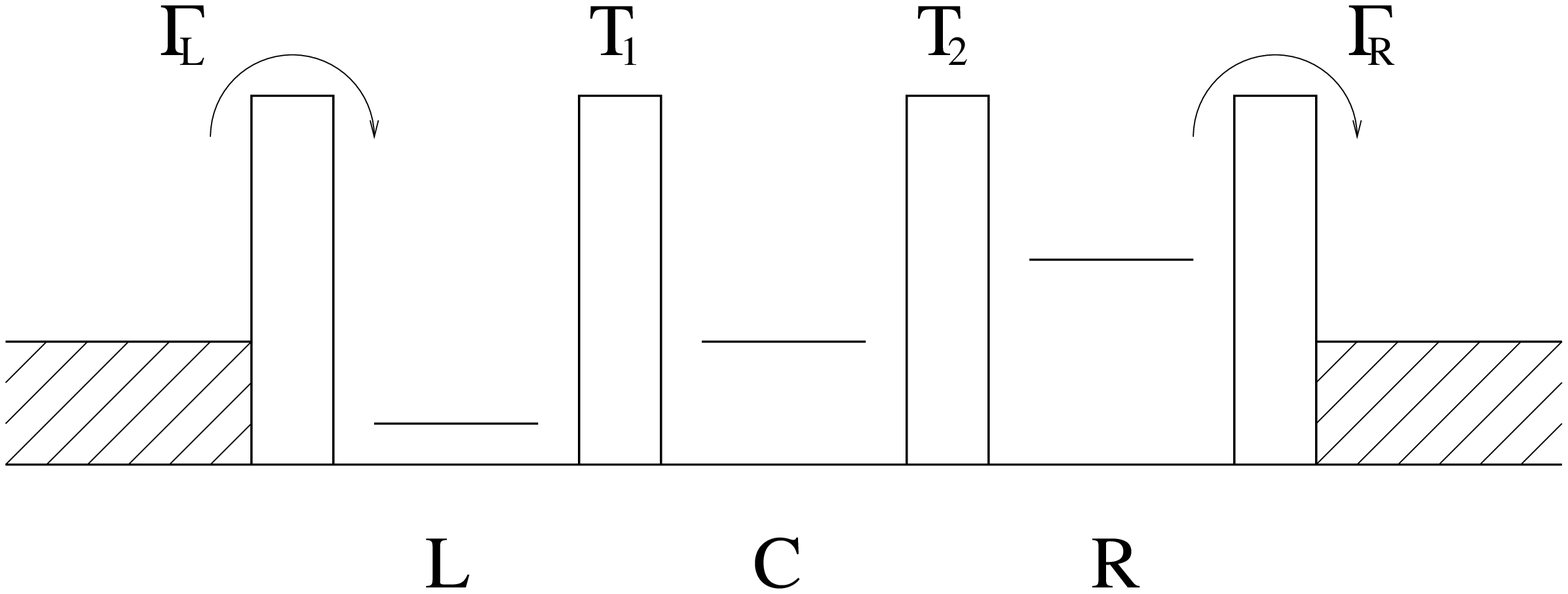}}
\end{picture}
\caption{Semiconductor structure consisting of three dots coupled via
tunnel barriers $T_1$ and $T_2$. The left and right dot are also coupled 
to electron reservoirs (leads).}
\label{fig_dots}
\end{figure}

The minimal structure for our scheme to work is shown in Fig. \ref{fig_dots}
and consists of three dots (L, C and R) coupled via time--varying tunnel
barriers. The left and the right dot are also coupled, via additional
tunnel barriers, to electron reservoirs.
Only the ground states take part in the process of charge transport and 
their energies are tuned, via gate voltages, so that $E_R>E_C>E_L$.
In the numerical calculations we will set $E_C=0$ and take $E_R=-E_L
\equiv E_0$. 
The energy splitting $E_0\equiv \hbar \omega_0$ will then
be taken as a scale for the energy and $1/\omega_0$ will be the time
unit.

We assume the charging energy of an additional electron in  the triple dot
so large that the system is effectively described by a four dimensional 
Hilbert space with basis $\{ |0\rangle, |L\rangle,|C\rangle,|R\rangle \}$. 
Here $|0\rangle$ is the "empty" state (no additional electron in the 
structure), and $|L\rangle$ ($|C\rangle, |R\rangle$) corresponds to an 
additional electron in the ground state of the left (center, right) dot.

We first discuss the effective one-particle problem of the isolated 
(no coupling to the leads) triple dot. The effective Hamiltonian is
\begin{eqnarray}
H(t)&=&E_L|L\rangle\langle L|+E_C|C\rangle\langle C|+E_R|R\rangle\langle R|+
\hbar T_1 (t) \cdot  \nonumber \\ 
& &\left[ |L\rangle\langle C| + |C\rangle\langle L| \right] +
\hbar T_2 (t) \left[ |C\rangle\langle R|+ |R\rangle\langle C| \right]~,
\label{ham}
\end{eqnarray}
with $E_{\alpha}$ ($\alpha=L,C,R$) the energy of the ground states of the 
different dots, and $T_1$, $T_2$ the tunneling constants, assumed to be
real (negative) for the sake of simplicity, between neighbouring dots.

To illustrate the principle of the proposed mechanism of charge 
transport, we neglect dephasing processes, such as the interaction with the
phonon bath, so that the evolution of the system is ideally coherent and 
corresponds to the wave function:
\begin{eqnarray}
|\psi (t)\rangle &=& c_L (t) \exp[-iE_Lt/\hbar] |L\rangle +
                 c_C (t) \exp[-iE_Ct/\hbar] |C\rangle + \nonumber \\
             &&  c_R (t) \exp[-iE_Rt/\hbar] |R\rangle
\end{eqnarray}
whose coefficients obey the Schr\"odinger equation 
\begin{mathletters}
\begin{eqnarray}
\dot{c}_L (t) &=& -i T_1 c_C(t) \exp[-i(E_C-E_L)t/\hbar]\\
\dot{c}_C (t) &=& -i T_1 c_L(t) \exp[-i(E_L-E_C)t/\hbar]\nonumber\\
              & & -i T_2 c_R(t) \exp[-i(E_R-E_C)t/\hbar]\\
\dot{c}_R (t) &=& -i T_2 c_C(t) \exp[-i(E_C-E_R)t/\hbar]~.
\end{eqnarray}
\label{sch}
\end{mathletters}
\begin{figure}[ht]
\setlength{\unitlength}{1in}
\begin{picture}(4,3)
\put(0,0){\epsfxsize 3 in \epsfbox{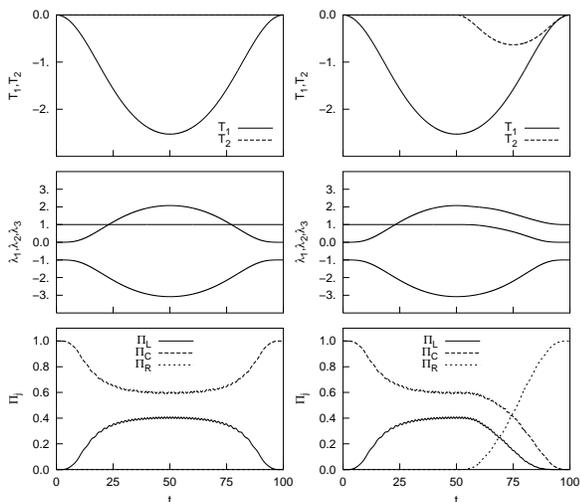}}
\end{picture}
\caption{Tunnel-coupling pulse sequence (top) and corresponding time
evolution of the energy eigenvalues (center) and populations (bottom). 
The populations $\Pi_{\alpha}=c^{*}_{\alpha} c_{\alpha}$ ($\alpha=L,C,R$)
are determined by numerically solving the Schr\"odinger equation Eq.
(\protect\ref{sch}).
The left column corresponds to the case in which only one tunnel-coupling
is nonzero ($T_2\equiv 0$), while for the right column both couplings are
pulsed.}
\label{fig_num2}
\end{figure}         
Consider first the ideal case in which the tunneling between neighbouring 
dots can be completely suppressed, i.e. the tunneling rates $T_1$ and $T_2$
can be tuned between zero and any arbitrary value $T_i\le 0$. 
Then, an appropriate tailoring of the time--varying tunnel rates between the 
dots results in pairs of level crossing--anticrossing, in profound analogy 
with the energy spectrum of atoms in static electric and magnetic fields 
\cite{windholz}. In the adiabatic limit \cite{messiah}, the state of the
system coincide with the instantaneous eigenstate of the Hamiltonian. 
Therefore it is possible to steer the wavefunction of the additional
electron by changing the parameters (tunnel couplings) of the system
Hamiltonian. 
In our case, adiabatic following corresponds to the transfer of charge
from one dot to its neighbour dot.
If the additional electron initially is in the center dot 
($c_{\alpha}(0)=\delta_{\alpha C}$), turning on the coupling $T_1$ induces
a mixing between the center and the left dot. As a result, the electron
spreads into the left dot (Fig. \ref{fig_num2}, bottom left) and the energy
levels of the coupled system (left and center dot) are pushed apart for
increasing $|T_1|$ (Fig. \ref{fig_num2}, center left). 

For $E_L < E_C < E_R $, a sufficiently large increase of $|T_1|$ results in 
a crossing of the energy level of the right dot with the higher energy level
of the coupled left-- and center--dot system. Clearly, if $T_2$ is kept at 
zero, a slow increase of $|T_1|$ and subsequent decrease to zero does
not produce any change of the state of the system, which follows adiabatically.
This is the situation described in the left column of Fig. \ref{fig_num2}. 
In this way no charge transport is produced. 

We  now show that a pulsed $T_2$ tunnel coupling  completely changes
the time evolution of the system, so that adiabatic following results in a
nonzero charge transport through the structure. This is the situation 
examined in the right column of Fig. \ref{fig_num2}. In fact, a pulsed $T_2$
can transform a level crossing into an anticrossing, as shown in Fig. 
\ref{fig_num2} (right column, center). If the $T_2$ pulse is shorter than 
the $T_1$ pulse and it is centered around the position of a crossing of the
unperturbed system (i.e. a crossing for $T_2\equiv 0$), only one of the 
level crossing will become an anticrossing. In this way the adiabatic 
following of the system results in the transfer of the additional electron 
from the initial dot to the neighbouring one as shown in Fig. \ref{fig_num2}.

\begin{figure}[th]
\setlength{\unitlength}{1in}
\begin{picture}(4,3)
\put(0,0){\epsfxsize 3in \epsfbox{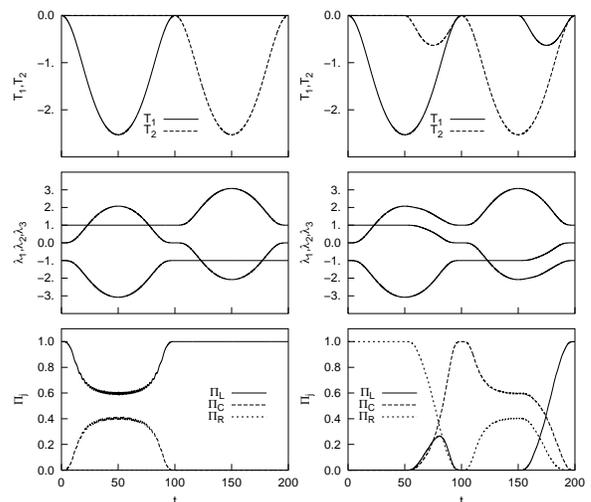}}
\end{picture}
\caption{Transfer from the right dot to the left one. The organization 
of the data sets is the same as for Fig. \protect\ref{fig_num2}.}
\label{fig_num3}
\end{figure}

The process can be iterated, and by exchanging the role of $T_1$ and 
$T_2$ the electron can be transferred adiabatically from one side 
of the structure (say, the right dot) to the other side (the left dot). 
The corresponding procedure is shown in Fig. \ref{fig_num3}. In fact, 
assuming the additional electron is initially localized in the right dot, 
it can be transfered to the center dot by the same sequence of 
tunnel--coupling pulses as above: a long $T_1$ pulse is applied, which alone 
would produce a pair of level crossing; a shorter $T_2$ pulse changes the
second level crossing into an  anticrossing, so that the electron is 
adiabatically transfered to the center dot. For the transfer from the
center dot to the left one the role of $T_1$ and $T_2$ are exchanged:
it is now $T_2$ which produces the pair of level crossings with $T_1$
playing the role of the control pulse, i.e. $T_1$ changes the last crossing
into an  anticrossing. In this way the adiabatic following results in 
the transfer of the additional electron to the left dot.

\begin{figure}[th]
\setlength{\unitlength}{1in}
\begin{picture}(4,2)
\put(0,0){\epsfxsize 3in \epsfbox{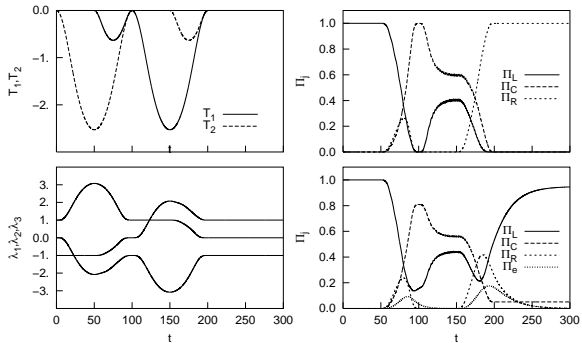}}
\end{picture}
\caption{Charge transport through tunnel-coupled quantum dots. 
Left column: the tunnel-coupling pulse sequence (top) and 
corresponding time evolution of the energies (bottom).
Right column: the evolution of the populations, as determined by
(\protect\ref{master}), including the coupling
to the leads ($\Gamma_L = 0.1$, $\Gamma_R = 0.05$, bottom), and for
a closed system ($\Gamma_L = \Gamma_R = 0$, top).}
\label{fig_master}
\end{figure}

We now include the coupling of the triple dot to the leads, and show that 
in this case the above mechanism allows to transfer electrons between two
reservoirs (the left and right leads) kept at the same chemical potential, 
see Fig. \ref{fig_dots}.  The corresponding tunneling rates are denoted 
by $\Gamma_L$ and $\Gamma_R$, respectively. The chemical potential of the 
electron reservoirs is tuned somewhere in between the energies of the left
and right dot, so that electrons can tunnel into the quantum-dots structure
only from the left lead and escape from the structure only  through the 
right barrier (see Fig. \ref{fig_dots}). 
For weak coupling to the leads, the dynamics of the coupled dots 
can be described by a master equation \cite{stoof,gurvitz,wegewijs}  
for the reduced density matrix $\rho$ of the dots, which in our case  reads

\begin{mathletters}
\begin{eqnarray}
\dot{\rho}_{LL} &=& i T_1 [\rho_{LC}-\rho_{CL}] + \Gamma_L \rho_{ee} \\
\dot{\rho}_{CC} &=& i T_2 [\rho_{CR}-\rho_{RC}] + 
                    i T_1 [\rho_{CL}-\rho_{LC}] \\
\dot{\rho}_{RR} &=& i T_2 [\rho_{RC}-\rho_{CR}] - \Gamma_R \rho_{RR}\\
\dot{\rho}_{ee}  &=& -\Gamma_L \rho_{ee} + \Gamma_R \rho_{RR} \\
\dot{\rho}_{LC} &=& i \omega_{CL}\rho_{LC}+i T_1 [\rho_{LL}-\rho_{CC}]
                    +i T_2 \rho_{LR} \\
\dot{\rho}_{CR} &=& i \omega_{RC} \rho_{CR} + i T_2  [\rho_{CC}-\rho_{RR}]
                    -i T_1 \rho_{LR} -\frac{1}{2}\Gamma_R \rho_{CR}\\
\dot{\rho}_{LR} &=& i \omega_{RL}\rho_{LR} - i T_1\rho_{CR}+ i T_2\rho_{LC}
                    -\frac{1}{2}\Gamma_R\rho_{LR}
\end{eqnarray}
\label{master}
\end{mathletters}
where $\omega_{\alpha,\beta}=(E_{\alpha}-E_{\beta})/\hbar$
($\alpha,\beta=C,L,R$) and ${\rho}_{ee}$ corresponds to the `empty' state.

The sequence of tunnel--couplings appropriate to transfer electrons from 
the left to the right reservoir is shown in Fig. \ref{fig_master} (top left),
together with the resulting energy eigenvalues (bottom left). We numerically 
solved the master equation (\ref{master}); results for the populations
$\Pi_{\alpha}=\rho_{\alpha\alpha}$ ($\alpha=e,L,C,R$) are reported in  
Fig. \ref{fig_master} (bottom right).
Results for a closed system, i.e. without coupling to the leads, are also 
reported for comparison (top right). In the latter case, the time evolution 
corresponds to the transfer of an additional electron from the left to the 
right dot.  In the case of an open system, the transfer of charge to the 
right dot is followed by a charge leakage to the right lead at a rate 
$\Gamma_R$. At the same time, charge flows from the left lead into the 
structure. In this way there is a net charge transport through the 
triple dot which after the tunnel--couplings sequence (including a "leakage
time" of the order of $1/{\rm min}(\Gamma_L,\Gamma_R)$) is returned to the
initial state with (almost) the whole charge in the left dot. Furthermore,
also the tunnel-couplings $T_1, T_2$ are  back to their initial zero value.

\begin{figure}[ht]
\setlength{\unitlength}{1in}
\begin{picture}(4,2)
\put(0,0){\epsfxsize 3in \epsfbox{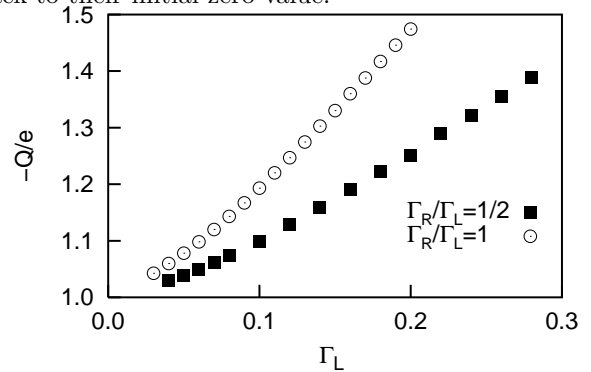}}
\end{picture}
\caption{Charge pumped through the triple dot in a cycle as a function of 
the tunnel rate $\Gamma_L$, at different values of $\Gamma_R/\Gamma_L$.}
\label{fig_charge}
\end{figure}

The total charge $Q$ pumped through the structure during a cycle, i.e.
during a time interval $[t_i:t_f]$ which includes the sequence of 
$T$-pulses and the leakage time, is
\begin{equation}
Q=-e\int_{t_i}^{t_f}\Gamma_L\rho_{ee}dt=-e\int_{t_i}^{t_f}
\Gamma_R\rho_{RR}dt~.
\label{charge}
\end{equation}
Note that Eq. (\ref{charge}) is valid only in integral form. In fact, the 
charge can pile up temporarily within the triple dot so that the instantaneous
value of $-e\Gamma_L\rho_{ee}$, i.e. the current flowing from the left 
lead into the dot--structure, is different from  $-e\Gamma_R\rho_{RR}$, the
current across the barrier between the right dot and the lead.

Results of numerical calculations for the charge $Q$ for different 
strengths of the coupling to the leads are reported in Fig. \ref{fig_charge}.
In the limit of weak coupling to the electron reservoirs, i.e. for small
$\Gamma_L$ and $\Gamma_R$, approximately one electron per cycle is
transferred from the left to the right lead. However, it should be noted
that a weaker coupling to the leads results in a slower electron pumping,
because of the longer leakage time.

The scheme for adiabatic transfer as described above is based on the 
existence of pairs of level crossings/anticrossings. In our structure,
a level {\em crossing} corresponds to the suppression of tunneling between
two neighbouring dots ($T_1$ or $T_2$ set to zero). However, it is in 
general not possible to completely suppress the tunneling between neighbouring 
dots, as this would correspond to infinitely high and/or thick tunnel 
barriers in real space. If the tunnel rates are kept at non--zero values
$T_i<0$ all the time, the previous degeneracies at the level crossings are
lifted and the crossings become anticrossings. This is consistent with the
fact that in a finite one--dimensional potential there are no level crossings
for a particle without internal structure \cite{LL}. 

We have repeated our calculations by adding a small but finite offset to the
tunnel rates, $T_i(t)\to T_0 + T_i(t)$. The time evolution of the system
essentially remains unchanged with the only exception being the previous 
level crossings that now turn into anti--crossings.
Although for small $T_0$ ($T_0=-10^{-6}$ in our calculation) the resulting 
level splitting is very tiny, the transfer mechanism across these points now
is Landau--Zener tunneling, whereas outside the `nearly crossings' the
dynamics remains adiabatic.  It should be noted that in the extreme case of
arbitrarily {\em slow} tuning of the $T_i(t)$, the Landau--Zener tunneling
becomes exponentially small and there is no transfer of charge at all 
any longer. 

Finally, we comment on parameter ranges relevant for a possible experimental
realization in coupled  semiconductor few--electron quantum dots. 
Experiments in double dots \cite{Fujetal} have demonstrated that 
three gate voltages can be used to tune the tunnel coupling $T$ and 
the ground state energies of two quantum dots, although  an independent  
manipulation of two couplings $T_1$ and $T_2$ in triple dots
is bound to be more difficult.
In our calculation, we assumed that the 
ground state energy difference $\hbar \omega_0$ between  two adjacent 
dots fulfills
$\hbar \omega_0 \ll U, \Delta$, where $U$ is the Coulomb charging energy and 
$\Delta$ the single particle level spacing within a single dot. Typical values
are $U, \Delta \sim 1$ meV in coupled lateral dots with a diameter of 
$\sim 200$ nm \cite{Fujetal} . 
Assuming $\hbar\omega_0\sim 0.1$meV,
we find that the typical operation frequency $\nu:=1/(t_f-t_i)$ of the pump 
is $\nu \sim 10^8$s$^{-1}$. The temperature smearing of the Fermi distribution 
is negligeable if $k_BT \ll \hbar \omega_0 \sim 1K$. For these parameters,
the inequality $h \nu,k_BT \ll \hbar \omega_0 \ll U, \Delta$ holds.

The described quantum pump is based on the possibility to steer the 
triple dot wavefunction by varying the tunnel couplings. In the intermediate
steps of a pumping cycle, the additional electron is indeed prepared in 
a superposition of states corresponding to neighbouring dots. As an 
additional test of the pure quantum nature of the proposed pumping
mechanism, we studied numerically the effects of an interdots relaxation 
rate $\gamma$ which transforms superposition of states into mixtures.
For the parameters mentioned above, $\nu$ is smaller or of the 
same order as rates $\gamma$ due to phonon emission. Fortunately, we verified 
that it takes relativey large relaxation rates $\gamma \gtrsim \omega_0$ 
for the charge pumped in a cycle to drop to zero. Further studies for the 
crossover to a completely incoherent transport regime are under way.

In conclusion, we have proposed a mechanism for charge pumping through 
triple quantum dots. Coupling to external reservoirs allows for adiabatic
pumping of electrons. The quantum state of the additional electron within 
the structure is steered by changing the tunnel couplings between
neighbouring dots. Appropriate tailoring of the time--dependent tunnel 
couplings allows the transfer of electrons from one dot to the other. 
Weak couplings to electron reservoirs then permits  to implement an 
electron pump.

\bibliographystyle{../base/prsty}

\end{document}